\documentclass[aps,pra,twocolumn, noshowpacs, groupedaddress,nopacs,floatfix]{revtex4}

\usepackage{graphicx}
\usepackage{amssymb}
\usepackage{amsmath}
\usepackage{times}

\begin{document}
\title{Atom-Light Interactions in Photonic Crystals}
\author{A. Goban$^{1,3,\dag}$, C.-L. Hung$^{1,3,\dag}$, S.-P. Yu$^{1,3,\dag}$, J. D. Hood$^{1,3,\dag}$, J. A. Muniz$^{1,3,\dag}$\footnotetext{\small $^{\dag}$ These authors contributed equally to this research.},
J. H. Lee$^{1,3}$,\\ M. J. Martin$^{1,3}$, A. C. McClung$^{1,3}$, K. S. Choi$^{4}$, D. E. Chang$^{5}$, O. Painter$^{2,3}$, and H. J. Kimble$^{1,3,\ast}$\footnotetext{\small $^{\ast}$ e-mail: hjkimble@caltech.edu.}}
\address{$^1$ Norman Bridge Laboratory of Physics 12-33}
\address{$^2$ Thomas J. Watson, Sr., Laboratory of Applied Physics 128-95}
\address{$^3$ Institute for Quantum Information and Matter, California Institute of Technology, Pasadena, CA 91125, USA}
\address{$^4$ Spin Convergence Research Center 39-1, Korea Institute of Science and Technology, Seoul 136-791, Korea}
\address{$^5$ ICFO - Institut de Ciencies Fotoniques, Mediterranean Technology Park, 08860 Castelldefels (Barcelona), Spain}

\begin{abstract}
The integration of nanophotonics and atomic physics has been a long-sought goal that would open new frontiers for optical physics. Here, we report the development of the first integrated optical circuit with a photonic crystal capable of both localizing and interfacing atoms with guided photons in the device. By aligning the optical bands of a photonic crystal waveguide (PCW) with selected atomic transitions, our platform provides new opportunities for novel quantum transport and many-body phenomena by way of photon-mediated atomic interactions along the PCW. From reflection spectra measured with average atom number $\bar{N} = 1.1 \pm 0.4$, we infer that atoms are localized within the PCW by Casimir-Polder and optical dipole forces. The fraction of single-atom radiative decay into the PCW is $\Gamma_{\rm 1D}/\Gamma^{\prime} \simeq (0.32 \pm 0.08)$, where $\Gamma_{\rm 1D}$ is the rate of emission into the guided mode and $\Gamma^{\prime}$ is the decay rate into all other channels. $\Gamma_{\rm 1D}/\Gamma^{\prime}$ is quoted without enhancement due to an external cavity and is unprecedented in all current atom-photon interfaces.
\end{abstract}
\maketitle

\fontfamily{ptm}\selectfont

\begin{figure}[t!]
\centering
\includegraphics[width=1\columnwidth]{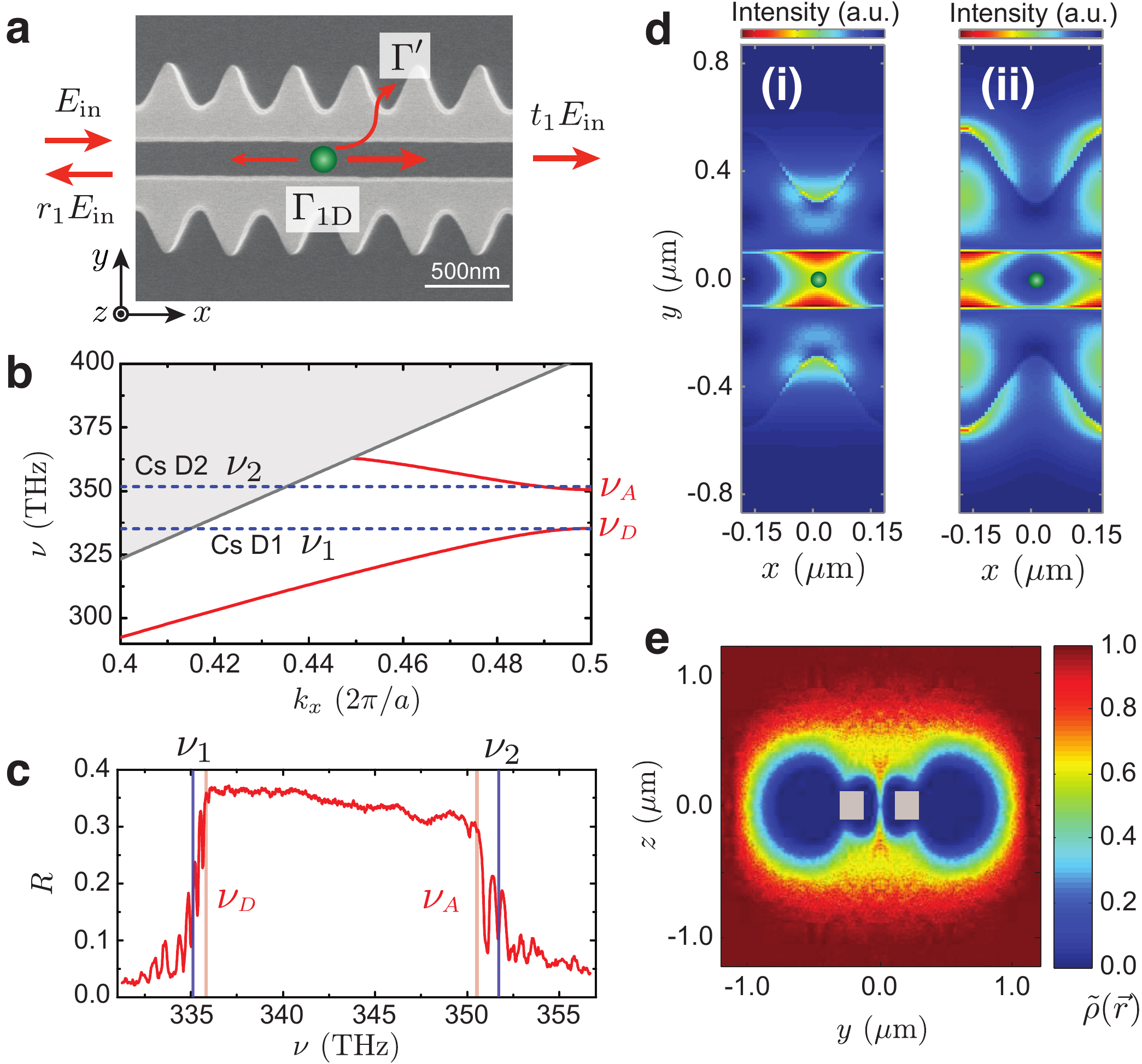}
\caption{Design and characterization of 1D photonic crystal waveguide. \textbf{a}, SEM image of the `alligator' photonic crystal waveguide (APCW) made from 200 nm thick (along $z$-axis) silicon nitride (SiN) \cite{yu2013}. Arrows indicate radiative processes of an atom (green circle) coupled to an incident electric field $E_{\rm in}$. \textbf{b}, Calculated band structure of fundamental TE-like modes $E_{1},~E_{2}$ (red solid lines), with dominant electric field polarized in the $y$-direction. The dashed lines mark the frequencies $\nu_1$ and $\nu_2$ of the cesium ${\rm D}_1,~{\rm D}_2$ lines, respectively. ($\nu_{A},~\nu_{D}$) mark near the band edge, respectively. The gray solid line marks the light-line. \textbf{c}, Measured reflection spectrum with fast fringe removed (see  Appendix B) around the band gap shown in \textbf{b}. (d) Intensity cross sections of (i), $E_2$ mode near $\nu_2$, and (ii), $E_1$ mode near $\nu_1$ within a unit cell calculated with the MPB software package~\cite{MPB} (see Appendix D). 
\textbf{e}, Simulated relative density $\tilde{\rho}(\vec{r})$ of atoms in the $x=0$ plane of \textbf{d}, (ii), with the optimal excitation of the blue-detuned $E_1$ mode at $k_x=\pi/a$ (see text).}
\label{fig1}%
\end{figure}

\begin{figure*}[t!]
\includegraphics[width=2\columnwidth]{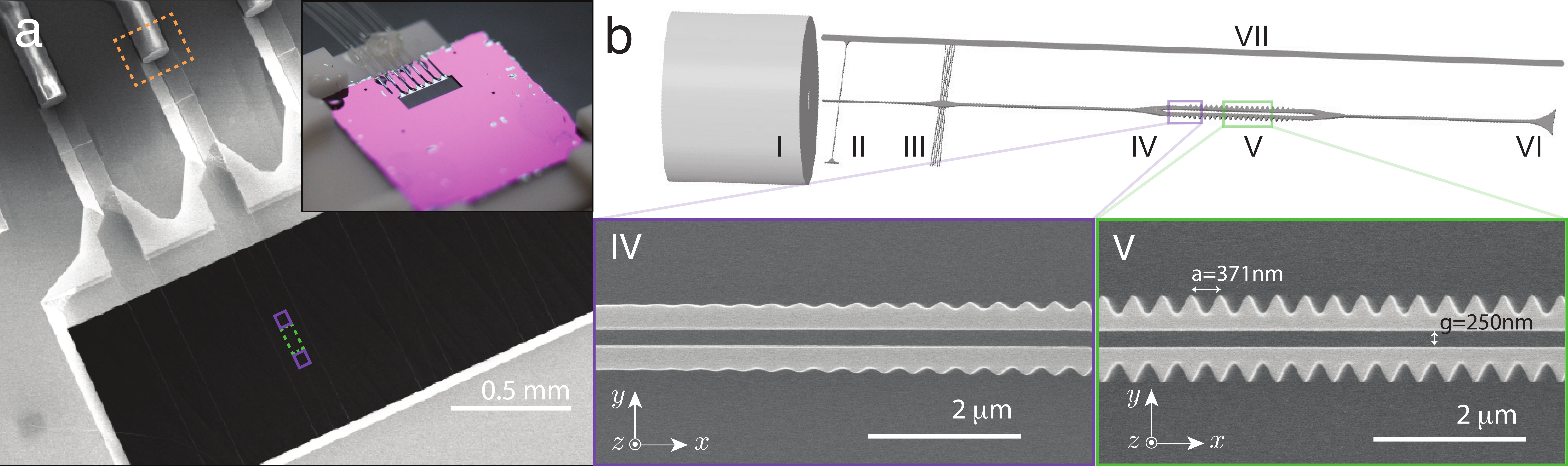}%
\caption{Overview of the integrated APCW device. \textbf{a} SEM image of the silicon chip showing an integrated optical fiber (orange box) coupled, via a SiN nanobeam waveguide, to the APCW region (purple and green boxes). The APCW is located within a 1 mm$\times$3 mm through window (black region without dielectrics) where free-space atoms and cooling lasers are introduced. Inset shows a picture of the chip and the optical fibers glued to a vacuum compatible holder. \textbf{b} Detailed schematic of the suspended SiN waveguide. Light enters the system via the optical fiber (I) butt-coupled \cite{cohen2013} to the free end of the waveguide (II) which is supported by a tether array (III). Near the center of the through window, the waveguide transitions into a double-nanobeam, followed by tapering (IV) and APCW (V) sections, and again tapers out to terminate into the substrate (VI). Two parallel rails are added symmetrically to support the structure (one rail is illustrated in VII). The insets, corresponding to the purple and green boxes in \textbf{a}, show SEM images of segments of the tapering (IV) and APCW (V) sections, respectively.}
\label{fig:Fabs}%
\end{figure*} 

Localizing arrays of atoms in photonic crystal waveguides with strong atom-photon interactions could provide new tools for quantum networks \cite{Kimble:2008,cirac97,duan04} and enable explorations of quantum many-body physics with engineered atom-photon interactions \cite{john1990,fan05,kien05,bhat06,dzsotjan10,Kien2008,Chang2007b,Zoubi2010,Chang2012,asboth_optomechanical_2008,gullans2012,chang2013,kurizki2013,shahmoom2013,chang2013c,chang2013b}. Bringing these scientific possibilities to fruition requires creation of an interdisciplinary `toolkit'  from atomic physics, quantum optics, and nanophotonics for the control, manipulation, and interaction of atoms and photons with a complexity and scalability not currently possible.
Here, we report advances that provide rudimentary capabilities for such a `toolkit' with atoms coupled to a photonic crystal waveguide (PCW). As illustrated in Fig.~\ref{fig1}, we have fabricated the first integrated optical circuit with a photonic crystal whose optical bands are aligned with atomic transitions for both trapping and interfacing atoms with guided photons \cite{hung2013,yu2013}. The quasi-1D PCW incorporates a novel design that has been fabricated in silicon nitride (SiN) \cite{cohen2013,yu2013} and integrated into an apparatus for delivering cold cesium atoms into the near field of the SiN structure. From a series of measurements of reflection spectra with $\bar{N} = 1.1 \pm 0.4$ atoms coupling to the PCW, we infer that the rate of single-atom radiative decay into the waveguide mode is $\Gamma_{\rm 1D} \simeq (0.32 \pm 0.08)\Gamma^{\prime}$, where $\Gamma^{\prime}$ is the radiative decay rate into all other channels. The corresponding single-atom reflectivity is $|r_1| \simeq 0.24$, representing an optical attenuation for one atom greater than $40\%$ \cite{hung2013,Chang2012}. For comparison, atoms trapped near the surface of a fused silica nanofiber exhibit  $\Gamma_{\rm 1D} \simeq (0.04 \pm 0.01)\Gamma^{\prime}$ \cite{Vetsch2010,Dawkins2011,Goban2012}, comparable to observations with atoms and molecules with strongly focused light \cite{tey2008,wrigge2008}. Here, $\Gamma_{\rm 1D}$ refers to the emission rate without enhancement  or inhibition due to an external cavity. By comparing with numerical simulations, our measurements suggest that atoms are guided to unit cells of the PCW by the combination of Casimir-Polder and optical dipole forces.

Important initial advances to integrate atomic systems and photonics have been made within the setting of cavity quantum electrodynamics (cQED) with atom-photon interactions enhanced in micro- and nanoscopic optical cavities \cite{vahala-review,vernooy98,Lev:2004,aoki2006,hinds2011,reichel2011,thompson2013} and waveguides \cite{Vetsch2010,Dawkins2011,Goban2012}. At a minimum, the further migration to photonic crystal structures should allow the relevant parameters associated with these paradigms to be pushed to their limits \cite{thompson2013} and greatly facilitate scaling. For example, modern lithographic processing can create nanoscopic dielectric waveguides and resonators with optical quality factors $Q>10^6$ and with efficient coupling among heterogeneous components \cite{barclay2007,Eichenfeld:2009,marshall2009,taguchi2011,bowers2011,lipson2013}.

A more intriguing possibility that has hardly been explored is the emergence of completely new paradigms beyond the cavity and waveguide models, which exploit the tremendous flexibility for modal and dispersion engineering of PCWs. For example, the ability to tune band edges near atomic transition frequencies can give rise to strongly enhanced optical interactions~\cite{Soljacic2002,Koenderink2006,lund08,hughes2010,Hoang2012}. This enables a single atom to exhibit nearly perfect emission into the guided modes~($\Gamma_{\rm 1D}\gg \Gamma'$) and to act as a highly reflective mirror~(e.g., reflection $|r_1| \gtrsim 0.95$ and transmission $|t_1| \lesssim 0.05$ for one atom \cite{hung2013}). The entanglement of photon transport with internal states of a single atom can form the basis for optical quantum information processing~\cite{Kimble:2008,cirac97,duan04} with on-chip quantum optical circuits. At the many-body level, the strong interplay between the optical response and large optical forces of many atomic ``mirrors'' can give rise to interesting opto-mechanical behavior, such as self-organization \cite{chang2013}.

Even more remarkable phenomena in PCWs arise when atomic frequencies can be tuned into photonic bandgaps, including the ability to control the range, strength, and functional form of optical interactions between atoms~\cite{john1990,kurizki2013,shahmoom2013,chang2013c}. For example, atoms trapped near otherwise perfect photonic crystal structures can act as dielectric defects that seed Òatom-inducedÓ cavities \cite{chang2013c} and thereby allow atomic excitations to be exchanged with proximal atoms \cite{kurizki2013}. The Òatom-inducedÓ cavities can be dynamically controlled with external lasers enabling the realization of nearly arbitrary long-range spin Hamiltonians and spatial interactions~(such as an effective Coulomb potential mediated by PCW photons)~\cite{chang2013c}, providing a novel tool for quantum simulation with cold atoms. Control over PCW dispersion is also expected to facilitate novel atomic traps based upon quantum vacuum forces~\cite{hung2013,chang2013b,Lamoreaux_2005}.
The prerequisite to all of these possibilities is a designable platform that allows the simultaneous alignment of optical bands for optical trapping and for interaction physics with atoms, which we demonstrate here for the first time.

Turning to our experiment, we begin with an SEM image of a small section of our 1D photonic crystal waveguide shown in Fig.\ref{fig1}a. The device consists of two parallel nanobeams with sinusoidal modulation at the outer edges (an `alligator' PCW or APCW). A challenge in the fabrication of PCWs for atomic physics is placement of the band edges near relevant atomic transition frequencies. Our APCW design facilitates this juxtaposition by fine tuning the gap between the parallel nanobeams and the amplitude of sinusoidal modulation in the APCW. Figure \ref{fig1}b shows the band structure of two fundamental transverse electric (TE-like) modes calculated based on the dimensions measured from SEM images as in Fig.\ref{fig1}a. The two blue dashed lines correspond to the transition frequencies of the Cs $({\rm D}_1,~{\rm D}_2)$ lines at $(\nu_1=335,\nu_2=351)$ THz [$(895,852)$nm] which straddle the band edge frequencies $(\nu_D,\nu_A)$ at $k_x=\pi/a$ for the lower (dielectric) and upper (air) TE bands, respectively. To validate these results, we measure the reflection spectrum $R(\nu)$ versus the input frequency $\nu$ for the actual device used in the reported experiments; see Fig.\ref{fig1}c. The large reflectivity ($R \sim 0.35$)
from $\nu_D$ to $\nu_A$ corresponds to the band gap for the APCW, while the vertical dashed lines mark $(\nu_1,\nu_2)$ in dispersive regions just outside the gap. Absent propagation loss to and from the APCW, we infer $R_{\rm gap}\simeq 0.99$ from measurements and numerical simulation.

Beyond band-edge placement, another requirement for realizing strong atom-light interactions in PCWs is to localize atoms in a region of high mode intensity within a unit cell. The use of two bands enables the separate engineering of the modes for trapping (lower band) and control of spontaneous emission (upper band).  The blue-detuned $E_1$ mode excited at $\nu_1$ in Fig.\ref{fig1}d (ii) can guide atoms into the center of the vacuum space near regions of large $|E_2|^2$, with then a field near $\nu_2$ serving as a probe mode.

The efficacy of this strategy is supported by trajectory calculations of free-space atoms surrounding the APCW (see Appendix D). As shown in Fig.\ref{fig1}e, atoms are guided from free space into the region of high $|E_2|^2$, resulting in a density approximately $30\%$ of the remote free-space density. For the simulations, the Casimir-Polder potential $U_{CP}(\vec{r})$ for the structure in Fig.\ref{fig1}a is computed numerically following Ref. \cite{hung2013}. The optical dipole potential is calculated using a guided mode $E_1(\vec{r})$ at $k_x=\pi/a$ with total power of 1~$\mu$W and 10~GHz blue-detuning from the $F=4 \leftrightarrow F^\prime=4$ transition frequency of the D$_1$ line. 

An overview of the integrated APCW device is presented in Fig. \ref{fig:Fabs}, and shows the optical pathways for excitation to and from the APCW, as well as the supporting structures of the SiN device to a silicon substrate.
The entire APCW contains $260$ unit cells with a lattice constant $a=371$nm, and is terminated on each end by a mode matching section of $40$ cells with tapered sinusoidal modulation and a transition section from a double- to a single-nanobeam waveguide. Input to and output from the device is achieved through an optical fiber butt-coupled to one of the single-nanobeam waveguides \cite{cohen2013}. The one-way efficiency for propagation from the APCW to the fiber mode is $T_{wf} \simeq 0.6$ (see Appendix A-C).

To integrate the device into a cold-atom apparatus, the silicon chip in the inset of Fig. \ref{fig:Fabs}(a) and its coupling optical fibers are mounted on a vacuum feedthrough with linear translation and rotation stages and inserted into a UHV chamber. Cesium atoms are delivered to the region surrounding the APCW by a three-stage process of transport and cooling. The resulting atomic cloud has a peak number density of $\rho_0 \simeq 2\times10^{10}$ /cm$^3$ at temperature $T \simeq 20~\mu$K measured via time-of-flight absorption imaging.

\begin{figure}[t]
\centering
\includegraphics[width=1\columnwidth]{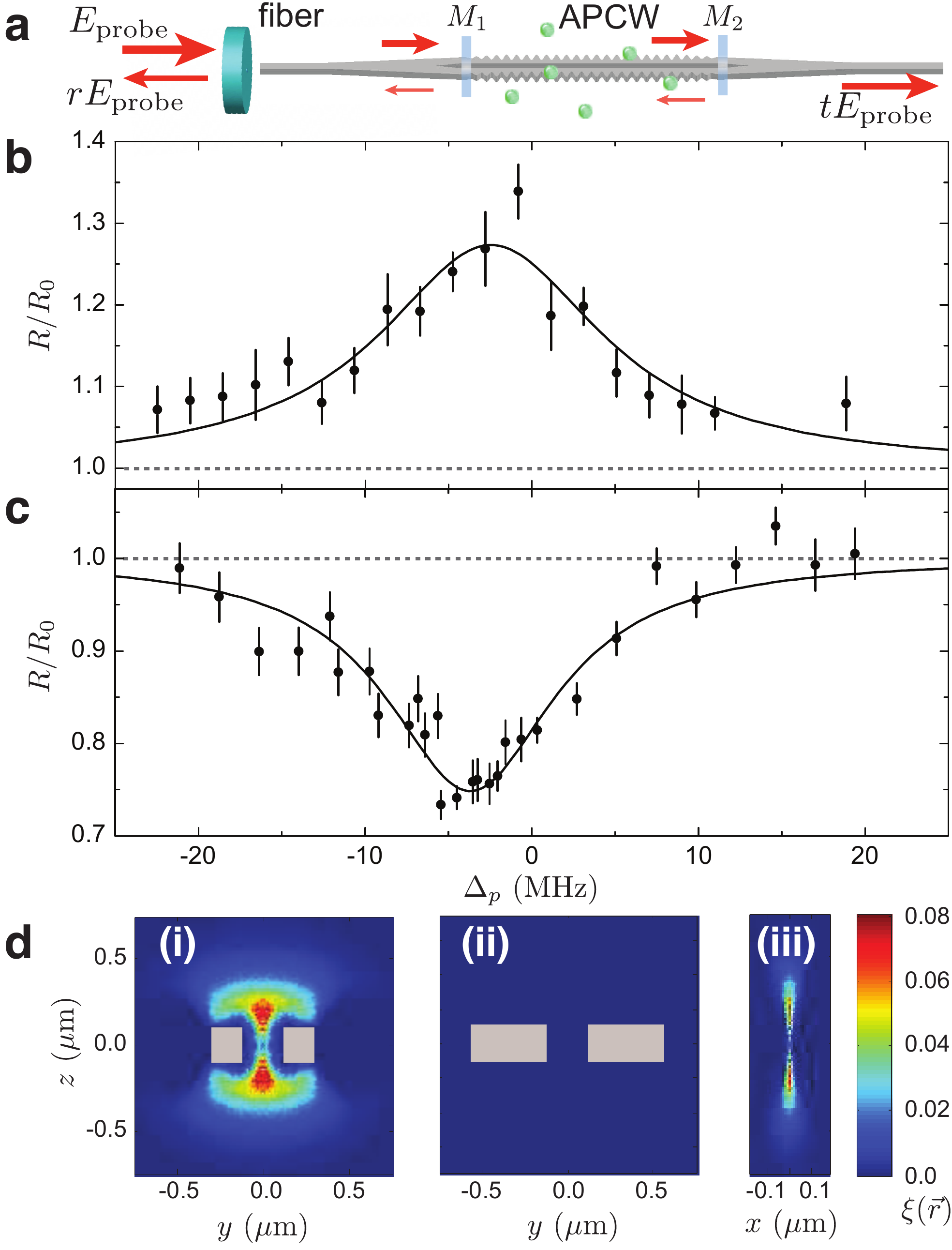}
\caption{Atom-light coupling in the photonic crystal waveguide section. \textbf{a}, Simplified schematic of a fiber-coupled photonic crystal waveguide for reflection measurements with atoms (green dots). Blue rectangles, marked as $M_1$ and $M_2$, illustrate the effective low-finesse $\mathcal{F}\simeq 2$ cavity formed by the tapered matching sections. \textbf{b}, \textbf{c} Measured reflection spectra (circles) with \textbf{b}, an on-resonant cavity and \textbf{c}, an off-resonant cavity in the APCW. Solid lines are Lorentzian fits with \textbf{b}, linewidth of 15.2$\pm$1.8 MHz, peak reflectivity $R/R_0=1.27\pm0.02$ and frequency shift $\Delta_0=-2.5\pm0.6$ MHz, and \textbf{c}, linewidth of 11.5$\pm$1.1 MHz, $R/R_0=0.75\pm0.01$ and $\Delta_0=-3.7\pm0.3$ MHz. Error bars for the data points reflect the statistical uncertainties.
\textbf{d}, Simulated $\xi (\vec{r}) = \tilde{\rho}(\vec{r}) \times \Gamma_{\rm 1D}(\vec{r})/\Gamma_{\rm 1D}(0)$ in the (i), $x=0$, (ii), $x=a/2$, and (iii), $y=0$ planes, with a guided potential of $m_F=0$ using the experimentally excited blue-detuned $E_1$ mode; see text and Appendix D for details. Masked areas in gray represent the APCW.}
\label{fig2}%
\end{figure}

We study atom-light interactions in the APCW by first shutting off the cooling laser, followed by a delay of $0.1$~ms, and then interrogating the APCW with atoms by sending a guided probe pulse $E_{\rm probe}$ of frequency $\nu_{p}$ in the $E_2$ mode, with typical power $\simeq 1$~pW; see Fig.\ref{fig2}a. Reflection spectra $R(\Delta_p)$ are recorded for 1 ms with a single-photon avalanche photodiode as a function of detuning $\Delta_{p}=\nu_p - \nu_{2a}$, where $\nu_{2a}$ is the free-space $F=4\leftrightarrow F^{\prime}=5$ transition frequency of the D$_2$ line. For $10$~ms following termination of the probe pulse, the atom cloud disperses, and then reference spectra $R_0(\Delta_p)$ are recorded for a second probe pulse for 1~ms. For all experiments, the guided mode $E_1$ is driven continuously with power $\simeq 0.6~\mu$W at $10$~GHz blue-detuning from the $F=4\leftrightarrow F^{\prime}=4$ transition of the D$_1$ line.

In the ideal case of a single atom in an infinite PCW, an incident probe beam would be reflected with amplitude coefficient $|r_1|=\Gamma_{\rm 1D}/(\Gamma_{\rm 1D}+\Gamma^{\prime})$,\cite{Chang2012} where $\Gamma_{\rm 1D}$ is the $F=4\leftrightarrow F^{\prime}=5$ transition of the Cs D$_2$ line. Strong spontaneous decay into the guided mode (and hence large $|r_1|$) results from the small area over which the guided mode is concentrated together with reduced group velocity for frequencies near a band edge. These two effects are incorporated into an effective mode area for an atom at location $\vec{r}$ within the APCW, namely $A_{\rm m} (\vec{r})= n_g \sigma \Gamma_0 /2\Gamma_{\rm 1D}(\vec{r})$, where $n_g\simeq2$ is the measured group index at $\nu_2$, $\sigma=1.4 \times 10^{-9}~$cm$^2$ is the free-space atom-photon cross section  for an unpolarized atom, and $\Gamma_0$ is the free-space rate of decay. For unpolarized atoms located at the center of a unit cell $\vec{r}=(0, 0, 0)$ in the APCW, we expect $A_{\rm m}(0) = 0.24~\mu$m$^2$, and hence $|r_1| \simeq 0.39$, where $\Gamma^\prime \simeq 0.9 \Gamma_0$ from numerical calculations \cite{hung2013}.

In the case of our actual device, the finite lengths of the taper sections lead to imperfect mode matching into the APCW near the band edge. As illustrated in Fig. \ref{fig2}(a), the matching sections partially reflect an incident probe pulse and form a low-finesse ($\mathcal{F} \simeq 2$) cavity around the APCW. These weak cavity resonances in the reflection spectrum near $\nu_2$ are shown in Fig.\ref{fig1}c (without atoms) and complicate the spectra taken with atoms relative to the ideal case, as discussed below.

In Figs.\ref{fig2}(b, c), we measure distinctive reflection spectra with cold atoms under two configurations of the APCW. 
Figure~\ref{fig2}(b) displays $R(\Delta_p)/R_0(\Delta_p)$ acquired near a resonance for the matching cavity. We observe an increased peak reflectivity $R/R_0 \simeq 1.27 \pm 0.02$. By comparison, with the matching cavity excited midway between two cavity resonances, we observe decreased reflectivity with a minimum $R/R_0 \simeq 0.75 \pm 0.01$ in Fig.\ref{fig2}b.

The reflection spectra in Figs.\ref{fig2}b, c represent strong evidence for atomic interactions with the guided mode $E_2$ of the APCW. Although the cavity formed by the matching sections has a low finesse, $R(\Delta_p)$ nevertheless depends on the cavity detuning as predicted, i.e., exhibiting approximately Lorentzian profiles for increased (decreased) $R(\Delta_p)/R_0(\Delta_p)$ for $ \nu_{2a}$ coincident with (mid-way between) the weak cavity resonances.

Moreover, our numerical simulations as in Fig.\ref{fig2}d suggest that the blue-detuned $E_1$ mode performs three functions: 1) it excludes atoms from the exterior of the APCW, as in Fig.\ref{fig2}d (ii), 2) it guides atoms into regions of large $E_2$ probe intensity near the center of the unit cells, as in Figs.\ref{fig2}d (i, iii), and 3) it expels atoms from the vicinity of other parts of the waveguide, e.g. the single-nanobeam regions in Fig.\ref{fig2}a, leaving only the APCW region with significant atom-field interactions. Together, these considerations enable us to infer quantitatively the single-atom emission rate absent reflections from the tapered sections, as we now describe.

\begin{figure}[t!hb]
\centering
\includegraphics[width=1\columnwidth]{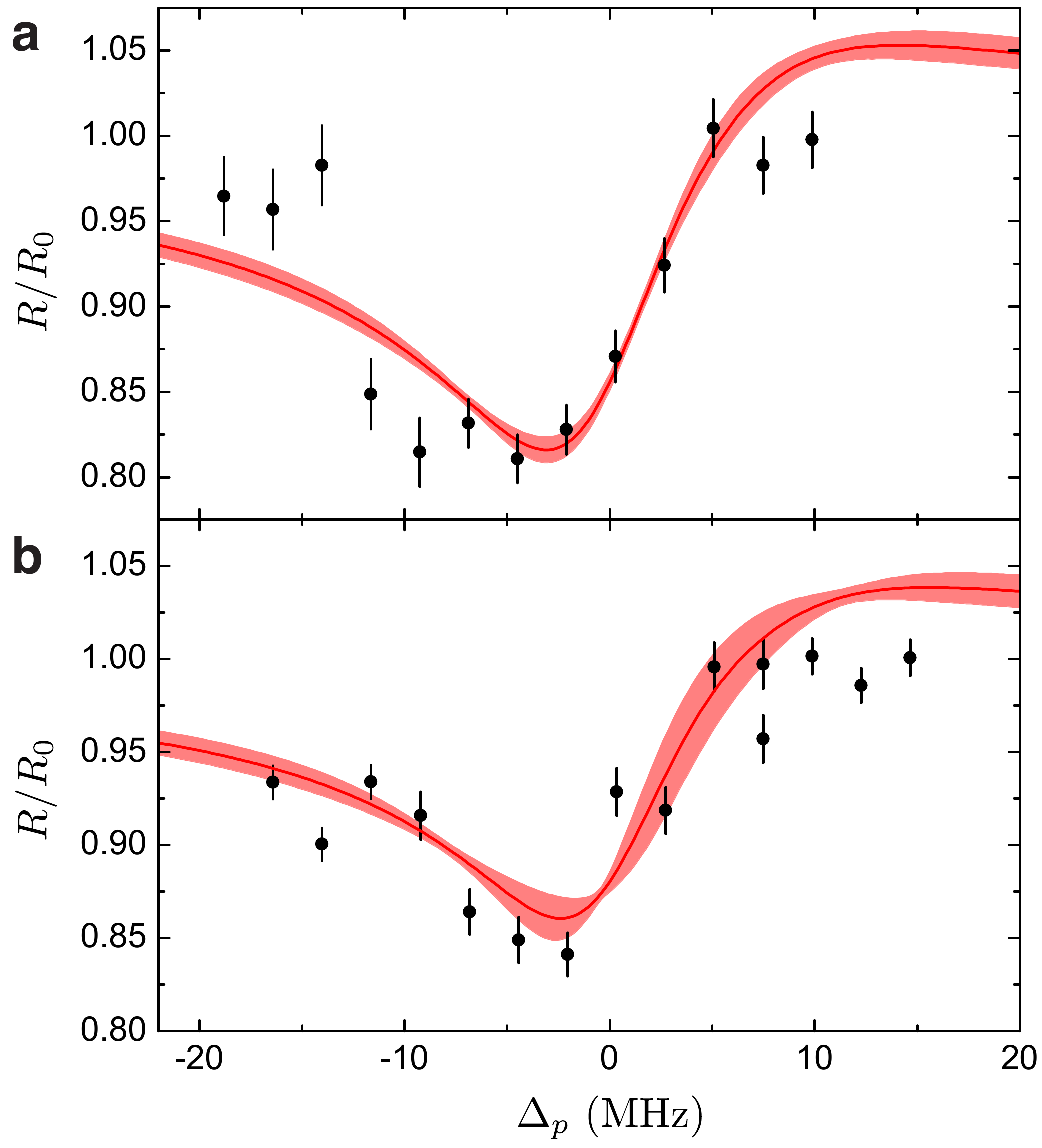}
\caption{Measured reflection spectra and theoretical fit for the APCW. Measured reflection spectra (circles) with free-space atomic cloud densities $\rho/\rho_0= 1$ \textbf{(a)}, and $0.75$ \textbf{(b)}. The full curves are fits with a model derived from transfer matrix calculations. Error bars for the data points reflect the statistical uncertainties. We deduce that $(\Gamma_{\rm 1D}/\Gamma^{\prime}, \bar{N}, \delta_0/\Gamma_0) \simeq$ $(0.35 \pm 0.1 , 1.0 \pm 0.1, 0.33\pm0.06)$ \textbf{(a)} and $(0.36 \pm 0.1, 0.76 \pm 0.13, 0.48\pm0.07)$ \textbf{(b)}. Here, $\rho_0 = 2\times 10^{10}/$cm$^3$; $\Gamma_0/2\pi = 5.2~$MHz (decay rate in free space). The shaded band gives the uncertainty arising from the position of the matching cavity (see Appendix F).}
\label{fig3}%
\end{figure}

To obtain quantitative information about atom-light coupling in the APCW region, we compare our measurements with a model based on transfer matrix calculations of the optical pathway to and from the APCW as illustrated in Fig.\ref{fig2}a. Details of the optical elements, including the coupling fiber, the supporting structures, and the APCW, are described in Ref.~\cite{yu2013}. Absent atoms, the optical characteristics of these various elements can be deduced from measurements of reflection spectra (c.f., Fig.\ref{fig1}c and Appendix C). With atoms, free parameters are the average atom number $\bar{N}$ within the APCW, the ratio $\Gamma_{\rm 1D}(\vec{0})/\Gamma^{\prime}$ for an atom at the center of the probe mode $E_2$, and the frequency shift  $\delta_0$ of the line center $\nu_{\rm 0}$ relative to free-space, $\delta_0=\nu_{\rm 0} - \nu_{2a}$. Here atoms are drawn from a Poisson distribution and placed randomly along the APCW.

Comparisons between measurements and our model for $R(\Delta_p)/R_0(\Delta_p)$ are given in Fig. \ref{fig3}. For these data, the weak cavity formed by the matching sections has a small detuning $\Delta_c \simeq 50~$GHz from the midpoint between two resonances (free spectral range $\sim$600 GHz). 
With atoms, the cavity detuning results in asymmetric, dispersive-like reflection spectra, which is captured by our model. From fits to the measured reflection spectra in Fig.~\ref{fig3}, we deduce that $\Gamma_{\rm 1D}/\Gamma^{\prime} \simeq 0.35 \pm 0.1 $ and $\bar{N}_0 \simeq 1 \pm 0.1$ for loading from a free-space cloud of density $\rho_0$. 

The inferred value of $\Gamma_{\rm 1D}$ allows us to determine $A_{\rm m}(\vec{r}_{\rm eff})$ for the atom-field interaction in our experiment, namely $A_{\rm m}(\vec{r}_{\rm eff}) \simeq 0.44~\mu$m$^2$ for an unpolarized atom. Together with the $E_2$ mode profile, the value of $A_{\rm m}$ suggests that atoms are distributed in narrow regions around $\vec{r}_{\rm eff} \simeq (0, 0, \pm 130)~$nm, which is consistent with our numerical simulations (Fig.\ref{fig2}d (i, iii)).

The large ratio $\Gamma_{\rm 1D}/\Gamma^{\prime} \simeq 0.35$ implies a single-atom reflectivity $|r_1| \simeq 0.26$, which is sufficient to give a nonlinear dependence of $R(\Delta_p)/R_0(\Delta_p)$ on the atom number $N$ observed in the measured spectra. We are thereby able to disambiguate the product $\bar{N}\times \Gamma_{\rm 1D}$ into separate parameters $\bar{N}, \Gamma_{\rm 1D}$ in fitting our model to measurement.

\begin{figure}
\centering
\includegraphics[width=1.0\columnwidth]{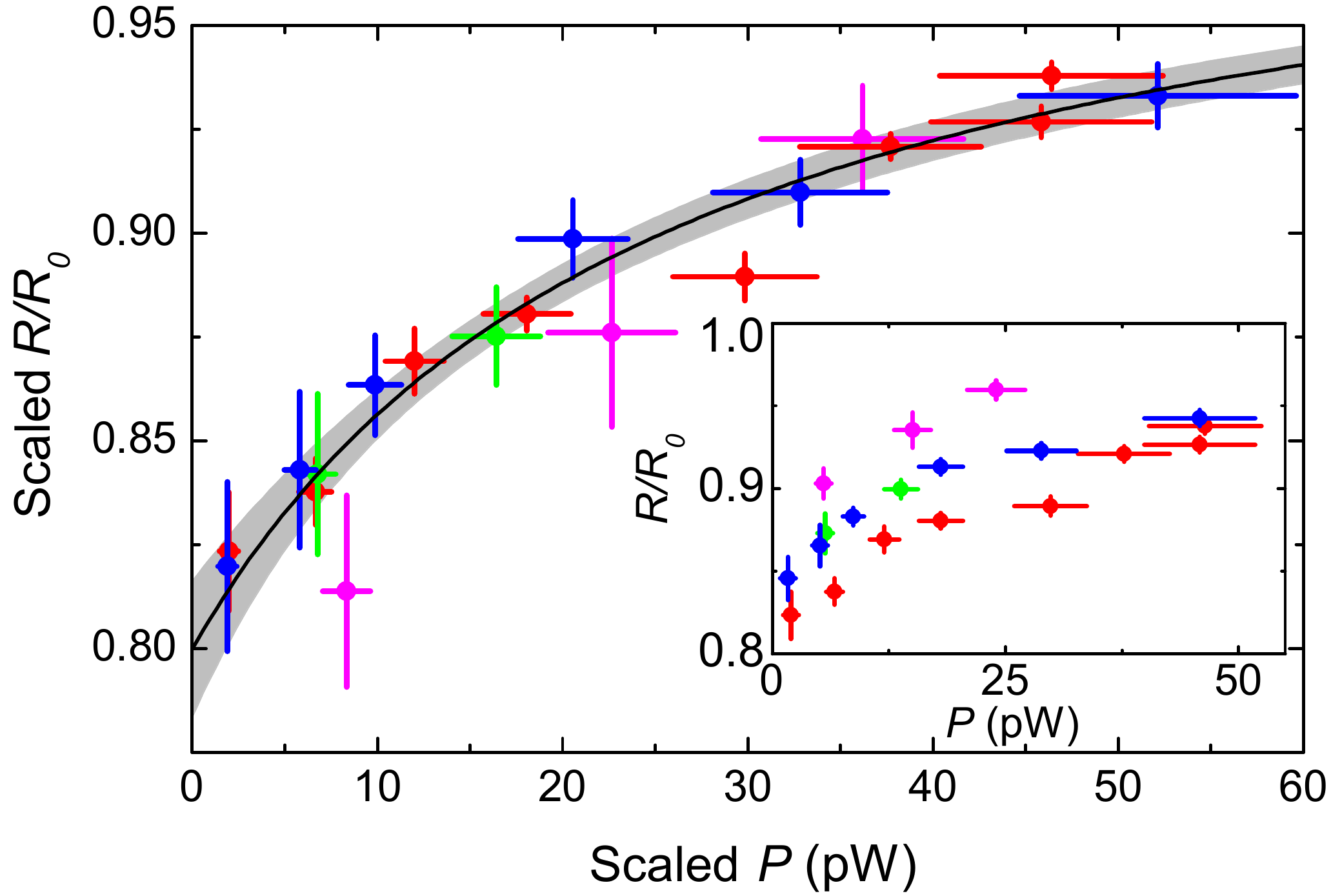}
\caption{Scaled reflection minimum as a function of averaged probe power inside the APCW. The saturation spectra (circles) are measured with four free-space densities $\rho/\rho_0$ = 1 (red), 0.81 (blue), 0.75 (green), 0.44 (magenta), and rescaled to a common free-space density $\rho_0\simeq2\times10^{10}$/ cm$^3$; see text. The inset shows the saturation data without scaling. An empirical fit (solid curve) gives a saturation power of 22.7$\pm$ 2.2 pW. Error bars for the data points reflect the statistical and
systematic uncertainties. The gray area shows 95\% confidence band.}
\label{fig4}
\end{figure}

To further investigate the nonlinear dependence of the reflection spectra on atom number, we measure reflection spectra for increasing values of probe power $P$ in the $E_2$ guided mode for various values of $\bar{N}$. Figure \ref{fig4} presents results for $R(\Delta_{p})/R_0(\Delta_{p})|_{\rm min}$ from a series of measurements as in Fig. \ref{fig3} for increasing $P$ with the plotted points corresponding to the minima of $R/R_0$ versus $\Delta_{p}$ for each spectrum. The inset of Fig. \ref{fig4} displays four sets of measurements showing that saturation of the atomic response (i.e.,  $R/R_0 \rightarrow 1$ with increasing $P$) requires higher power as the density is increased (i.e., increasing $\bar{N}$).

The observed saturation behavior can be scaled into a common curve using the dependence of cooperative atomic emission on atom number $\bar{N}$. We assume that $R(P) = R(P/P_{\rm sat})$ and that the saturation power $P_{\rm sat}$ depends on the average total decay rate as $P_{\rm sat} \propto (\Gamma' + \bar{N}\Gamma_{\rm 1D})^2$ with $\Gamma_{\rm 1D}/\Gamma^\prime =0.35 \pm 0.1$, as determined from our measurements in Fig. \ref{fig3} with $P\rightarrow 0$. We rescale the probe power (horizontal axis) for each density in the inset of Fig.~\ref{fig4} to a common density $\rho_0$. Likewise, $R(P)/R_0$ (vertical axis) is rescaled using the density dependence derived from our transfer matrix model (Appendix F), with $\bar{N}\propto \rho$. The approximate convergence of the data to a common curve in Fig. \ref{fig4} supports our rudimentary understanding of the underlying atom-field interactions in the APCW, including that the observed line shapes for the data in Fig. \ref{fig4} taken at higher power (not shown) are predominately homogeneously broadened.

To estimate the saturation power in the APCW, we adopt an empirical form for the saturation behavior $R(P, \bar{N})/R_0 = \exp\{- \gamma(\bar{N}) /[1+P/P_{\rm sat}(\bar{N})]\}$. From the fit in Fig.~\ref{fig4}, we determine $P_{\rm sat}(\bar{N}_0)=22.7\pm2.2$ pW and $\gamma(\bar{N}_0)=0.22\pm0.01$. Combining with the measured effective mode area, we find the saturation intensity $I_{\rm sat} = P_{\rm sat}/A_{\rm m} \simeq 5.2~$mW/cm$^2$, close to the expected value $I_{\rm s0} \times(\Gamma' + \bar{N}_0 \Gamma_{\rm 1D})^2/\Gamma_0^2 \simeq 4.0~$mW/cm$^2$, where $I_{\rm s0} = 2.7~$mW/cm$^2$ is the free-space saturation intensity.

In conclusion, we have realized a novel APCW device for interfacing atoms and photons. The measured coupling rate $\Gamma_{\rm 1D}$ (quoted absent Purcell enhancement and inhibition due to an external cavity) is unprecedented in all current atom-photon interfaces, whether for atoms trapped near a nanofiber \cite{Vetsch2010, Dawkins2011, Goban2012}, one atom in free-space \cite{tey2008}, or a single molecule on a surface \cite{wrigge2008}. For example, in Ref. \cite{thompson2013} a drop in transmission $\simeq 0.02$ is observed for single atoms trapped outside a photonic crystal cavity. In our work without trapping, we observe a dip in reflection $\simeq 0.25$ for $\bar{N} \simeq 1$ since atoms are channeled to near the peak of the probe mode in the center of unit cells with stronger interactions. Further improvements to the APCW include active tuning of the band edge to near an atomic resonance to achieve an increase $\gtrsim 50$ fold in $\Gamma_{\rm{1D}}$ \cite{hung2013}, as well as tuning to place the atomic resonance within the band gap to induce long-range atom-atom interactions \cite{john1990,kurizki2013,shahmoom2013,chang2013c}. By optimizing the power and detuning of the $E_1$ trap mode, we should be able to achieve stable atomic trapping and ground state cooling \cite{hung2013,thompson2012,kaufman2012}. By applying continuous onsite cooling to $N \gg 1$ atoms, we expect to create a 1D atomic lattice with single atoms trapped in unit cells along the APCW, thus opening new opportunities for studying novel quantum transport and many-body phenomena \cite{fan05,kien05,bhat06,dzsotjan10,Kien2008,Chang2007b,Zoubi2010,Chang2012,asboth_optomechanical_2008,gullans2012,chang2013,kurizki2013,shahmoom2013,chang2013c}.

\textit{Acknowledgements} We gratefully acknowledge the contributions of D. Alton, J. Cohen, D. Ding, P. Forn-Diaz, S. Meenehan, R. Norte, and M. Pototschnig. Funding is provided by the IQIM, an NSF Physics Frontiers Center with support of the Moore Foundation, the DARPA ORCHID program, the AFOSR QuMPASS MURI, the DoD NSSEFF program (HJK), and NSF PHY-1205729 (HJK).
AG is supported by the Nakajima Foundation. SPY and JAM acknowledge support from the International Fulbright Science and Technology Award.
The research of KSC is supported by the KIST institutional program.
DEC acknowledges funding from Fundaci\'{o} Privada Cellex Barcelona.

\appendix
\section{Design principle}
\begin{figure}[t!]
\includegraphics[width=1\columnwidth]{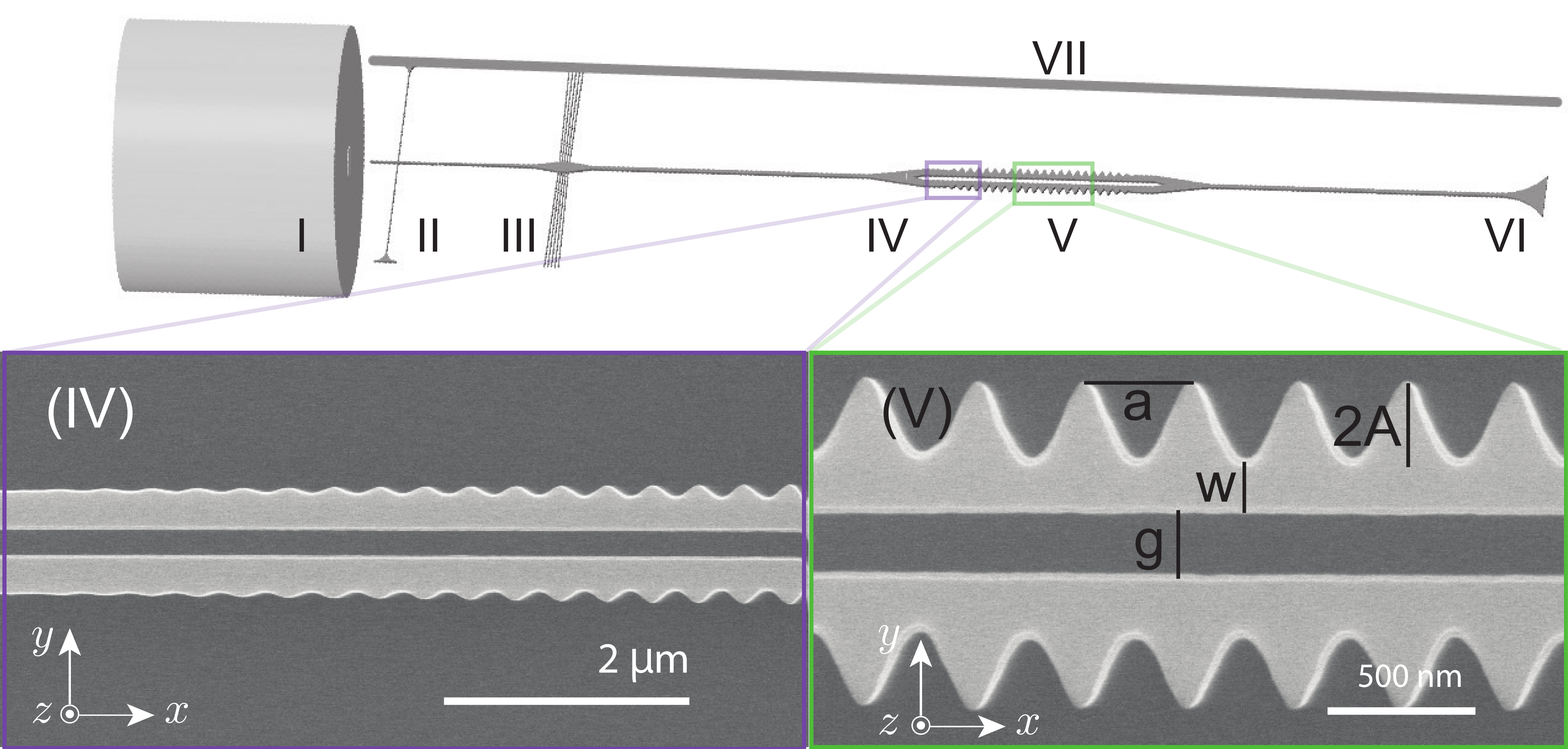}%
\caption{Schematic image of the suspended SiN device. Probe light enters the system via optical fiber (I) and couples into the free end of a tapered SiN waveguide (II) \cite{cohen2013}. The guided light transmits through the supporting tether array (III). At the center of the window, the waveguide is tapered into a double-nanobeam, and then tapered (IV) into the nominal APCW section (V) where atom-light interaction occurs. At the end of the window, the waveguide is tapered out and terminated into the substrate (VI). Two support rails are added symmetrically parallel to the APCW for structural integrity (one rail is illustrated in VII). The insets at the bottom are SEM images of (IV) the tapered section, and (V) the ``alligator" photonic crystal.}
\label{fig:Fab}%
\end{figure}

An ``alligator" photonic crystal waveguide (APCW) is designed on a chip in order to observe strong atom-light interactions, as illustrated in Fig. \ref{fig:Fab}.  The APCW interacts with a cloud of cesium atoms trapped in a magneto-optical trap (MOT) which is centered on the photonic crystal.  The APCW (see inset V of Fig.\ref{fig:Fab}) consists of two parallel nanobeams with sinusoidal corrugations on the outer edges \cite{yu2013}.  
The atoms are guided into the center of the two nanobeams by a scheme which takes advantages of the TE-like modes (y-polarized) near the band edges~\cite{MarkosBook}.  Their highly symmetric mode profiles near the dielectric and air bands at frequencies $(\nu_D, \nu_A)$ and the proximity to the resonant frequencies $(\nu_1, \nu_2)$ of the cesium D$_1$, D$_2$  lines allow us to create strong dipole potentials with small optical power ($<10~\mu$W) in the $E_1$ mode, while achieving large atom-photon coupling in the $E_2$ mode. See Figs.1(d) (i, ii) for calculated mode profiles. 

The corrugations in the APCW are used rather than the more traditional holes because the corrugation amplitude can be patterned more accurately than hole radii, resulting in more accurate alignment of the bandgaps and better adiabatic tapers. The waveguide is made from 200 nm thick stoichiometric silicon nitride (SiN) with index $n=2.0$ \cite{yu2013}.  
The degrees of freedom for the APCW are the gap $g$, lattice constant $a$, width $w$ (inner-edge to center of peaks), and tooth amplitude $A$.  The APCW has 260 cells, a gap 250 nm, a width 173 nm, and tooth amplitude 132 nm. The photonic crystal is tapered on both sides into an unpatterned (translationally invariant) double-nanobeam. The length and profile of the tapering determines the reflections from the edges of the APCW. The taper used here  has 40 cells, and is carefully designed such that the bandgap symmetrically opens about 873 nm, which is between the cesium D$_1$ and D$_2$ lines. The profile and the unpatterned double waveguide width are chosen to minimize reflections from the APCW edge.

In order to provide optical access for the trapping and cooling laser beams, the silicon nitride (SiN) waveguides are suspended across a 1 mm long 3 mm wide through-window on the chip, shown in Fig.2(a), and the APCW (the inset (D,E) of Fig.\ref{fig:Fab}) is positioned at the center. The suspended waveguide extends beyond the window into a V-groove etched into the Si substrate and then reduces in width for efficient coupling to a conventional optical fiber ~\cite{cohen2013} (sections I-II of Fig.\ref{fig:Fab}). The silicon anisotropic etch that forms the window also forms the V-shaped groove, which serves to center the fiber to the waveguide. The far end of the waveguide is extended to a fan shape and terminated into the substrate to minimize reflection (VI).

\begin{figure}[t!b]
\centering
\includegraphics[width=1\columnwidth]{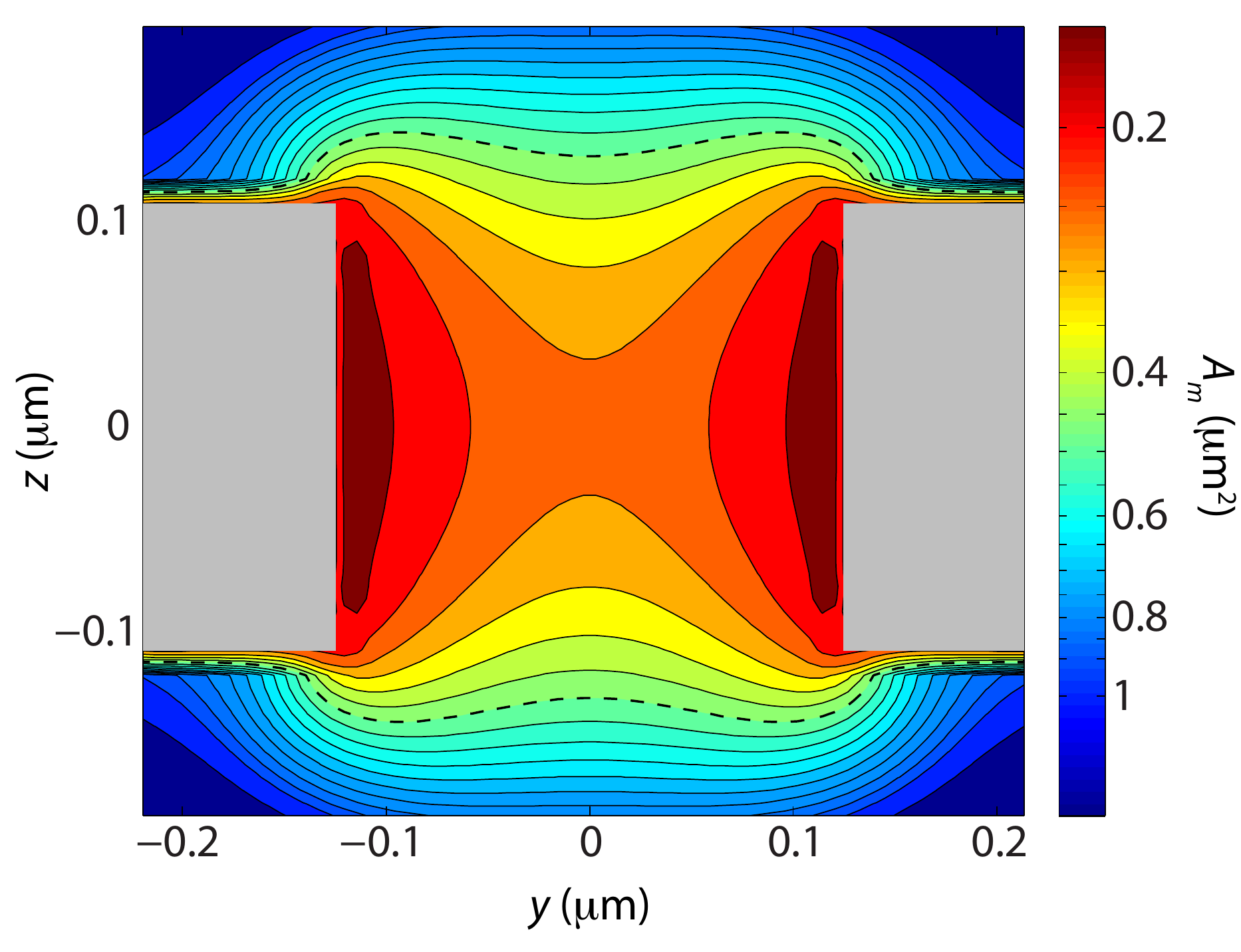}
\caption{Calculated effective mode area $A_{\rm m}(\vec{r}_{\rm A})$ for an atom at $\vec{r}_{\rm A}=(0,y,z)$ at the central $x=0$ plane of a unit cell. Masked areas in gray represent dielectric regions of the APCW. The dashed line represents $A_{\rm m}(\vec{r}_{\rm eff})\sim0.44~\mu$m$^2$ which we extract from measurements shown in Fig.3}
\label{fig:SM_Amode}%
\end{figure}

To characterize the strength of atom-photon coupling, we calculate the effective mode area  for an atom at $\vec{r}_{\rm A}$
\begin{equation}
A_{\rm m}(\vec{r}_{\rm A}) = \frac{ \int \epsilon (\vec{r'}) |E_2 (\vec{r'})|^2 d^3 r'}{a \epsilon(\vec{r}_{\rm A}) |E_2(\vec{r}_{\rm A})|^2},
\end{equation}
where $a$ is a lattice constant of 371 nm and the integration runs over the space occupied by a unit cell. 
Figure~\ref{fig:SM_Amode} shows $A_{\rm m}(\vec{r}_{\rm A})$ plotted at the central $x=0$ plane. For single atoms channeled to the center of  unit cells, we have $A_{\rm m}(0) = 0.24~\mu$m$^2$. 

To estimate atomic emission rate into the guided mode, we use $\Gamma_{\rm 1D}(\vec{r}) = \Gamma_0 n_g \sigma /2A_{\rm m}(\vec{r}_{\rm A}) $, where $\Gamma_0$ is the atomic decay rate in free space, $\sigma$ the radiative cross section, and $n_g$ the group index at $\nu_2$. When the band edge frequency $\nu_A$ is placed near $\nu_2$, we expect $n_g \gg 1$ due to slow light effect \cite{hung2013}; $n_g \simeq 2$ for the current device. For atoms guided to the center of unit cells and driven by the $F=4, m_F=0 \leftrightarrow F' = 5, m_{F'}=0$~transition, $\sigma \simeq 5/9 \times 3\lambda^2/2\pi$, where $\lambda=852~$nm is the free-space wavelength of the Cs D$_2$ line, with then $\Gamma_{\rm 1D}/ \Gamma_0 \simeq 0.4 n_g$; for unpolarized atoms, we calculate an averaged $\Gamma_{\rm 1D} / \Gamma_0 \simeq 0.29 n_g$.

\section{Device characterization}
\begin{figure}[t!]
\centering
\includegraphics[width=1\columnwidth]{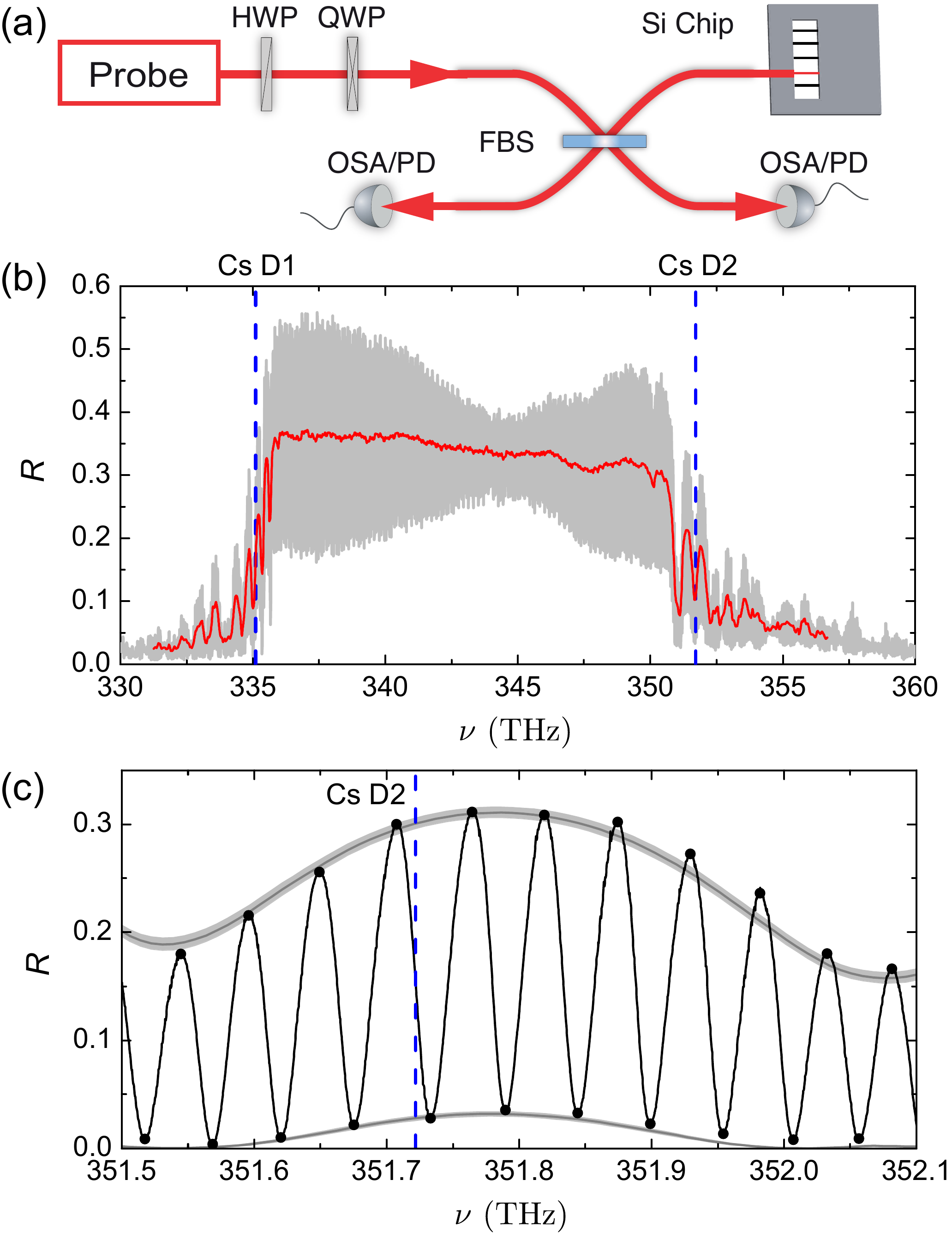}
\caption{(a) Schematic of the setup for device characterization. FBS: fiber beam splitter with~$T=50\%$ and $R=50\%$. HWP: half waveplate. QWP: quarter waveplate. OSA: optical spectrum analyzer. PD: photodetector. (b) Measured reflection spectrum near the bandgap. Gray and red curve show the original data and smoothed data, respectively. Blue dashed lines correspond to the Cs D$_1$ and D$_2$ resonances.
$($c) Measured reflection spectrum near the Cs D$_2$ line (black curve) and envelope fit (gray curve).}%
\label{fig:SMosa}%
\end{figure}

Figure \ref{fig:SMosa} (a) shows a schematic of the setup for device characterization. The reflection spectrum near the bandgap is measured with a broadband light source and an optical spectrum analyzer, by taking the ratio of spectral power between reference beam and reflected beam and considering the loss of each optical element.
Finer reflection spectra around the D$_1$ and D$_2$ lines are measured by scanning the frequency of narrow bandwidth diode lasers. The polarization in the device is aligned to the TE-like mode by observing the polarization dependent scattering from the first tether in the coupler, or equivalently by maximizing the reflected signal, since the transverse magnetic (TM-like) mode bandgap is located at a higher frequency.

Figure \ref{fig:SMosa} (b) shows the measured reflection spectra around the bandgap. The gray curve shows the original data which has a fast fringe (free spectral range$\sim50$ GHz) resulting from the (parasitic) etalon formed from the cleaved fiber end-face and the first matching taper. The red curve shows the smoothed reflection spectrum which approximately represents the response of the APCW without the influence of the reflection from the fiber end-face.

\section{Device model}
\begin{figure}[t!]
\centering
\includegraphics[width=1\columnwidth]{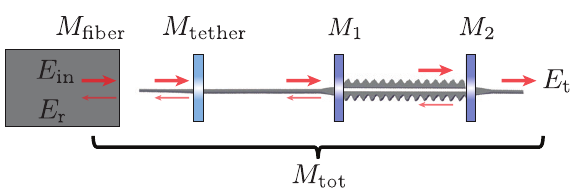}%
\caption{Schematic of a 1D transfer matrix model of the device. Forward- and backward propagating waves in the structure are coupled to each other through partial reflection from the fiber end-face, tether, and matching mirrors, and also evolve under propagation losses. Their effect on the total reflected and transmitted fields ($E_{\rm r}, E_{\rm t}$) can be modeled through a set of four independent transfer matrices $M_{\rm fiber}$, $M_{\rm tether}$, $M_1$, and $M_2$ (depicted by the blue rectangles), which together determine the transfer matrix $M_{\rm tot}$ of the entire system.}%
\label{fig:devicemodel}%
\end{figure}
Due to imperfection of the adiabatic tapering, the terminal regions of the APCW form a low finesse cavity. When atoms couple to light in the APCW, the reflected spectrum depends on the position of the frequency $\nu_2$ on the cavity fringe and can take on dispersive line shapes. Further complicating the picture is the presence of a fast $\sim$50 GHz fringe due to the parasitic etalons formed by the fiber end-face, tethers, and the APCW band edges near $\nu_D,\nu_A$.  In order to fit the reflected atomic signals, a full model which incorporates all of these elements is developed  using the transfer matrix method\cite{MarkosBook, Fan2002apl} as show in Fig.\ref{fig:devicemodel}. A transfer matrix represents each element, and the reflection, transmission, and loss coefficients are determined by both experiment and FDTD simulations~\cite{Lumerical}.

The light is coupled into the device by matching the modes of a 780HP single mode fiber (mode field diameter 5 $\mu$m) to a 130 nm wide rectangular SiN waveguide (I-II in Fig.\ref{fig:Fab} (b) ). The fiber end-face reflects 3.8$\%$ power due to the index mismatch. A 90 nm wide tether anchors the coupling waveguide 5 $\mu$m from the free (input) end of the waveguide, which has a theoretical transmission of 87$\%$ and reflection 0.8$\%$. The waveguide width tapers to 200 nm over 300 ~$\mu$m in order to better confine the light to the dielectric, and then the light propagates through the region of the support rails (III in Fig.\ref{fig:Fab} (b) ) to the APCW at the center of the window. Our numerical simulations show that the taper, support rails, and guide should have negligible loss and reflection. The loss in these sections was measured to be 22$\%$ loss per mm  for a similar device. The details of loss mechanism is currently under investigation.
The total loss from the fiber face to the waveguide can be estimated experimentally by measuring the reflected signal for frequencies within the bandgap, assuming the reflectivity of the APCW from numerical simulations is $\sim99\%$. By fitting our model to the envelope of the reflection spectrum inside the bandgap, we obtain the overall transmission efficiency from the internal face of the fiber to the input of the APCW to be $T_t\simeq0.60$, including propagation losses in the nanobeam.

The tapers of the APCW (represented by the matching mirrors in Fig.\ref{fig:devicemodel}) also reflect near the band edge. There is also loss inside the APCW due to fabrication disorder and absorption.  Near the Cs D$_2$ line, as shown in the Fig.\ref{fig:SMosa} $($c), the fitted spectrum (solid gray line) yields the reflection and transmission of the matching sections of the APCW, ($R_{\rm pc}=0.28, T_{\rm pc}=0.72$), the slope of the reflection of the APCW $dR_{\rm pc}/d\lambda=0.082$/nm, and the one-way transmission inside the APCW, $T_{\rm APCW}=0.40\pm0.02$.

\section{Simulations of relative density} 
\begin{figure*}[t!b]
\centering
\includegraphics[width=2\columnwidth]{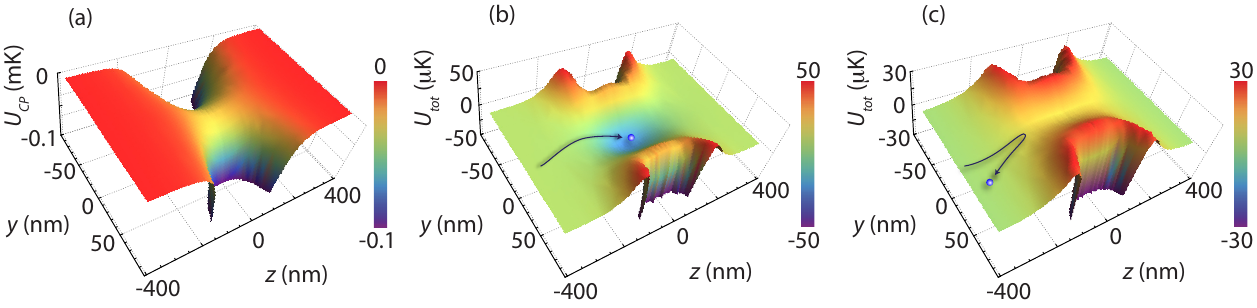}
\caption{Guiding potential for Cs $F = 4$ hyperfine ground state. (a) Casimir-Polder potential $U_{\rm CP}(\vec{r})$ and (b, c) total potential $U_{\rm tot}(\vec{r}) = U_{\rm CP}(\vec{r}) + U_{\rm dipole}(\vec{r})$ are plotted at $\vec{r} = (0, y, z)$ at the central $x = 0$ plane of a unit cell. The dipole potential $U_{\rm dipole}(\vec{r})$ is calculated for $m_F=0$ state, using the $E_1$ mode at (b) $k_{A, x} = \pi/a$ with total power of $1~\mu$W and (c) $k_{1, x} = 0.99 k_{A,x}$ with total power of $0.6~\mu$W, and both with $10~$GHz blue-detuning from $F=4 \leftrightarrow F^\prime =4 $ transition frequency of the D$_1$ line. The curved arrow in (b)[(c)] illustrates a characteristic trajectory of an atom (solid circle) guided into the trap center [reflected off the bump] in the total potential between the gap.}%
\label{fig:SMpot}%
\end{figure*}

To estimate a relative atomic density near the APCW with a guiding potential, we calculate a relative density $\tilde{\rho}(\vec{r})=\rho(\vec{r})/\rho_0$, where $\rho_0$ is a free-space cloud density, with a Monte Carlo simulation of $5\times10^6$ trajectories of thermal atoms with a temperature of 20 $\mu$K~\cite{Sague2007}. 
For the simulations, the Casimir-Polder potential $U_{\rm CP}(\vec{r})$ for the APCW is computed numerically following Ref. \cite{hung2013}, with an example cut shown in Fig.\ref{fig:SMpot}(a). The dipole potential $U_{\rm dipole}(\vec{r})$ of the blue-detuned guided mode $E_1(\vec{r})$ near $\nu_1$ is calculated by using the mode function obtained with MIT Photonic-Band package \cite{MPB}.
Trajectories are obtained by numerically solving the equation of motion with force of $\vec{F}=-\vec{\nabla} U_{\rm tot}(\vec{r})=-\vec{\nabla} (U_{\rm CP}(\vec{r})+U_{\rm dipole}(\vec{r}))$.
The relative density is inferred from atomic flux crossing each grid.
Note that velocity-dependent forces of polarization-gradient cooling (PGC) are not included in the simulations.

In Figs.1(e) and \ref{fig:SMpot} (b), we use a guided mode $E_1(\vec{r})$ at the band edge $k_{A, x}=\pi/a$ at $\nu_D$ with total power of 1 ${\rm \mu W}$ and 10 GHz blue-detuning from $F=4\leftrightarrow F'=4$ transition frequency of D$_1$ line, which has zero intensity at the center of unit cells of the APCW. 
Thus, atoms are channeled into unit cells by the combination of Casimir-Polder and optical dipole force. The relative density at the center of unit cells is $\tilde{\rho}(0)\sim0.3$.

In Figs.2(d) and \ref{fig:SMpot} (c), we use experimentally excited $E_1(\vec{r})$ with $k_{1,x} $ near $\nu_1$ (Cs D$_1$ line) which is $\sim1\%$ below the band edge $k_{A, x}$. Due to the small deviation from the band edge, the intensity of the mode field $E_1(\vec{r})$ near $\nu_1$ has a small bump at the center of unit cells, which for blue detuning, prevents atoms from being channeled to the center.
In addition, the intensity of $E_1(\vec{r})$ has a longitudinal component $E_{x}$ along the propagation direction, which is $\pi/2$ out-of-phase with the transverse components. As a consequence, the polarization of $E_1(\vec{r})$ is elliptical everywhere except for the central $y=0$ plane of unit cells due to TE symmetry. The resulting guiding potential has vector shifts due to the ellipticity of  $E_1(\vec{r})$, which lead to $m_F$ dependent guiding potentials. Since the center of guiding channel for $|m_F|\neq0$ states moves from $y=0$ to $|y|>0$ due to a stronger fictitious magnetic field (along z) closer to the structure, the guiding efficiency from the combination of Casimir-Polder force is reduced as $|m_F|$ increases. 

Figure 2(d) displays $\xi(\vec{r})=\tilde{\rho}(\vec{r})\times\Gamma_{\rm 1D}(\vec{r})/\Gamma_{\rm 1D}(0)$, where $\tilde{\rho}(\vec{r})$ is simulated relative atomic density with a guided potential for $m_F=0$, and $\Gamma_{\rm 1D}(0)$ is a decay rate into the APCW at the center of unit cells. Due to the small deviation from the band edge and the resulting potential `bump', atoms are localized around $>100$ nm from the center of unit cells.

We note that, in the case of Fig.\ref{fig:SMpot}(c), adding to $U_{\rm dipole}$ the dipole potential from a weak red-detuned $E_2(\vec{r})$ mode can help overcome the potential bump in the gap center and can create a stable trapping condition. This scheme is currently under investigation and will be presented elsewhere.

\section{Experimental procedure}

To prevent cesium contamination of the APCW due to the background vapor pressure, our vacuum system consists of a source chamber and a science chamber, connected via a differential pumping stage. The source chamber runs in a standard MOT loaded from the background Cs vapor. From the source MOT, a pulsed push beam extracts a flux of cold atoms that is slow enough to be recaptured in a science MOT located in the UHV region~\cite{Wohllebena2001}. We load a science MOT for $1$ s and compress it for a duration of 50 ms~\cite{DePue2001}. We then obtain a cloud of cold atoms with a peak density of $\sim1\times10^{11}$ /cm$^3$ and a temperature of $40~\mu$K about 1 cm away from the APCW of a silicon chip.

In order to transport cold atoms near the APCW, we cool atoms in the moving frame toward the APCW by abruptly changing the center of magnetic quadrupole field~\cite{Shang1991}. After shutting off cooling beams, a cloud of atoms freely propagates toward the APCW.  By turning on additional MOT beams at the time atoms fly near the APCW, we cool and recapture propagating atoms with an efficiency of $\sim40\%$. After applying PGC, we obtain a cloud of cold atoms with a peak number density of $\rho_0 \sim2\times10^{10}$ /cm$^3$, spatially overlapped with the APCW. 

\begin{figure}[t!]
\centering
\includegraphics[width=1\columnwidth]{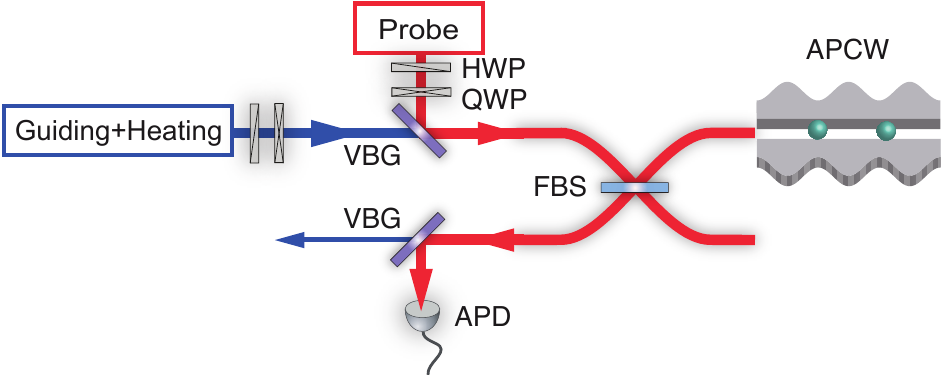}
\caption{Schematic of the setup for reflection measurements. VBG: volume Bragg grating. FBS:
fiber beam splitter with~$T=99\%$ and $R=1\%$. HWP: half waveplate. QWP: quarter waveplate. APD: avalanche photodetector.}%
\label{fig:SMexp}%
\end{figure}

An overview of reflection measurements in our experiment with cold atoms near the APCW is given in Fig.\ref{fig:SMexp}. The blue-detuned guiding beam with power $\sim0.6\mu$W and detuning of +10 GHz from D$_1$ ($F=4\leftrightarrow F'=4$) at $k_{1, x} = 0.99 k_{A,x}$ is on throughout the experiment. 
The guiding beam is combined with the probe pulse by a volume Bragg grating (VBG) and then couples to the APCW via a fiber beam splitter with~$T=99\%$ and $R=1\%$. The reflected probe signal from the APCW is efficiently picked up by the fiber beam splitter through the transmission path. An additional VBG at the output reflects the return probe beam, which allows us to measure the probe pulses with the guiding beam on. A pair of $\lambda/2$ and $\lambda/4$ waveplates in each path is used for aligning the polarization to only excite the TE-like mode.

To maximize the signal-to-noise ratio, we perform reflection measurements with atoms after aligning the bottom of the fast fringe to the Cs D$_2$ line, where the probe field is maximized inside the parasitic cavity formed by the fiber end-face and the APCW region; see Fig.\ref{fig:SMmodel}(a). The alignment of the fast fringe can be tuned by sending additional few $\mu$W heating beam to heat up the device and adjust the optical path length between the fiber end-face and the APCW region. The heating beam runs at frequency $\nu > \nu_D$ inside the band gap and is $\simeq 5~$nm detuned from D$_1$ line, thus does not interfere with atomic-light interaction in the APCW region. At the bottom of the fast fringe, for the experiment presented in Fig.3b, measured reflection without atoms is $R_0=0.3\%$; in Fig.3c, $R_0=3\%$.

\section{Model of reflection spectrum of atoms}

\begin{figure}[t!]
\centering
\includegraphics[width=1\columnwidth]{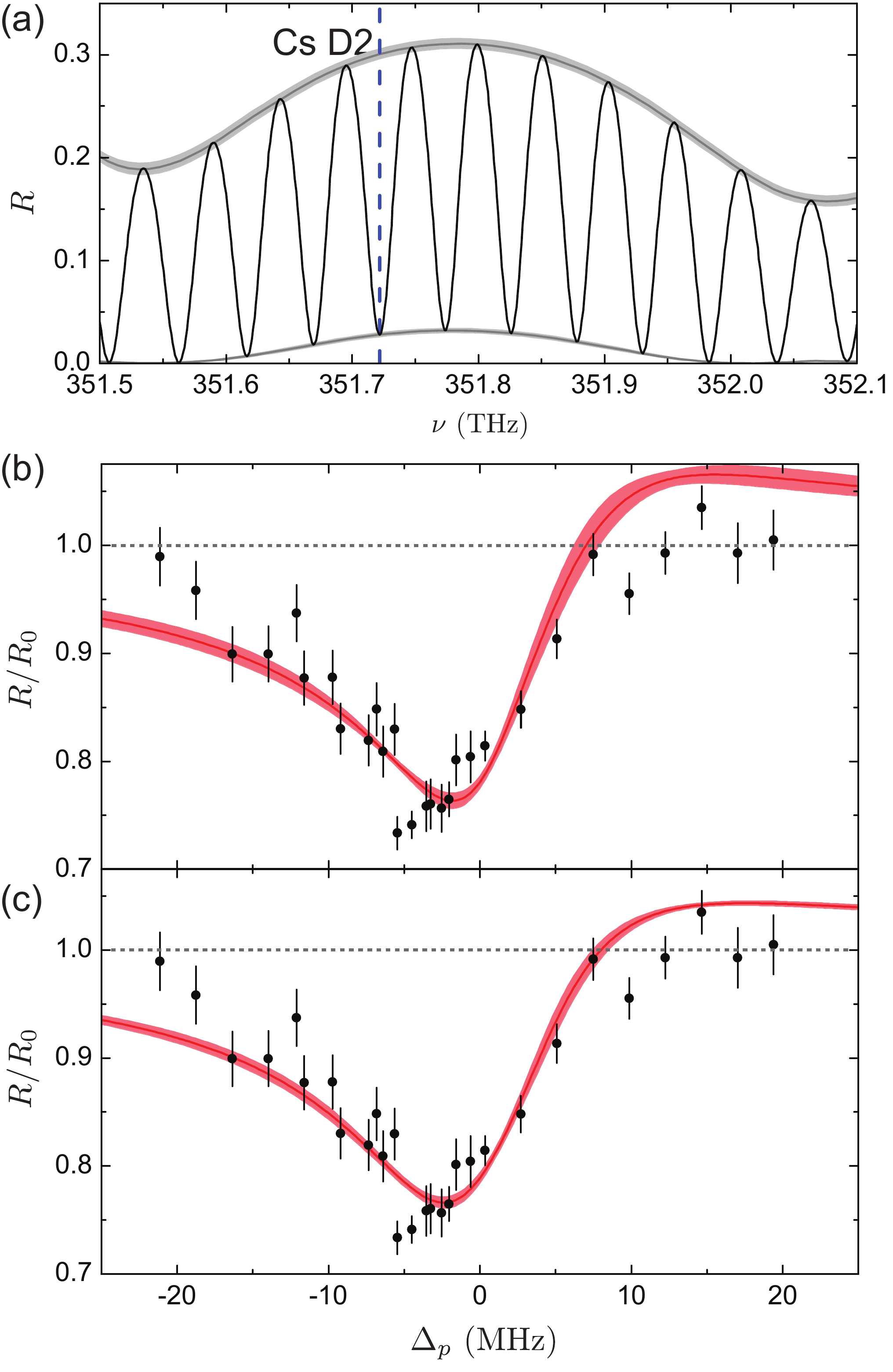}
\caption{(a) Simulated reflection spectrum of the optical pathway to and from the APCW derived from the transfer matrix calculation. Blue dashed line shows the Cs D$_2$ line. (b, c) Measured reflection spectra (circles) with free-space density of $\rho_0\sim2\times10^{10}$ /cm$^3$. These are the data same as Fig.2$($b). The full curves are fits with (b) uniform-absorption model (i) and (c) losses localized to the matching mirrors (ii). From the fits, we deduce  (b) $(\Gamma_{1D}/\Gamma', \bar{N}, \delta_0/\Gamma_0)\simeq(0.31\pm0.05, 1.5 \pm0.2, 0.56\pm0.06)$ and  $($c) ($\Gamma_{1D}/\Gamma', \bar{N}, \delta_0/\Gamma_0 )\simeq(0.41\pm0.04, 0.9 \pm0.1, 0.25\pm0.06)$. The shaded band gives uncertain arising from the position of the matching cavity.}%
\label{fig:SMmodel}%
\end{figure}

Reflection spectra of guided atoms are obtained by including transfer matrices for atoms~\cite{Deutsch1995} in the device model described in Section III.
Guided atoms inside the APCW are drawn from a Poisson distribution with mean atom number $\bar{N}$ and randomly placed at the center of unit cells along the APCW.
Each of the two matching sections that terminate the APCW partially reflects light near the band edge, as depicted in Fig.\ref{fig:devicemodel}. Together the matching sections form a cavity around the APCW (denoted by $M_1, M_2$ in Fig.\ref{fig:devicemodel}), whose cavity length has a frequency dependence.
We incorporate the uncertainty of the location of the matching mirrors and resulting cavity relative to the APCW into our model. This uncertainty gives rise to a variation in the reflection spectra from our model, which is given by the thickness of the lines shown in Fig.\ref{fig:SMmodel}. 

The wave vector of probe frequency $k_{{\rm p}, x}$ is $\simeq 2\%$ from the band edge at $k_{A, x}=\pi/a$ as shown in Fig.1b. In the case of more than one atom coupled to the APCW,  the mismatch between $k_{A,x}$ and  $k_{{\rm p}, x}$ causes dephasing since the accumulated phase between atoms is not an integer multiple of $\pi$. Due to the moderate one-way transmission of $T_{\rm APCW}\simeq0.40$ inside the APCW as described in Section III, the cooperative effect of two atoms is sensitive to how loss occurs inside of the APCW. Here, we consider two limiting cases: (i) uniform absorption along the APCW, and (ii) loss at the first matching mirror. 

In the case of (i), due to uniform loss along the APCW, only nearby atoms interact equally with the probe field.
Thus, cooperative effects survive despite of the mismatch of $k_{{\rm p}, x}$ and $k_{A, x}$.
On the other hand, in the case of (ii), all atoms contribute equally and cooperative effects are washed out due to the phase mismatch for propagation with $k_{{\rm p}, x}$ as compared to $k_{A, x}$ for the unit cell. Given our current limited knowledge of the microscopic details of the loss mechanisms within the APCW, the models (i, ii) provide a means to estimate the uncertainties in our inferences of $\Gamma_{1D}/\Gamma'$ and average atomic number $\bar{N}$ based upon comparisons of our data with the models.

For the cases shown in Fig.\ref{fig:SMmodel}, the model fits lead to the following: (i) $\Gamma_{1D}/\Gamma'=0.31\pm0.05$ and average atom number $\bar{N}\simeq 1.5 \pm0.2$ in (b), and (ii) $\Gamma_{1D}/\Gamma'=0.41\pm0.04$ and $\bar{N}\simeq 0.9 \pm0.1$ in (c). 

From four sets of data as in Fig.\ref{fig:SMmodel} taken for comparable atomic densities, we make fits based upon the two models (i) and (ii). We average    results for parameters determined from the fits to arrive at the values quoted, namely $\Gamma_{1D}/\Gamma'=0.32\pm0.08$, $\bar{N}=1.1\pm0.4$ and $\delta_0/\Gamma_0=0.13\pm0.27$.


\end{document}